\documentclass[aps,prd,preprint,a4paper,showpacs,nofootinbib,superscriptaddress]{revtex4-1}
\pdfoutput=1
\usepackage{bm}
\usepackage{indentfirst}
\usepackage{amsmath}
\usepackage{graphicx}
\usepackage{float}
\usepackage{amssymb}
\usepackage{subfigure}
\usepackage{amssymb}
\usepackage{hyperref}
\usepackage{epstopdf}
\usepackage[utf8]{inputenc}
\hypersetup{
    colorlinks=true,
    linkcolor=red,
    citecolor=blue,
}
\usepackage{color}
\usepackage[section]{placeins}%可以禁止图片或者表格浮动超过\section 的范围
\begin{document}
\baselineskip=0.5 cm

\title{Acoustic black hole in Schwarzschild spacetime: quasi-normal modes, analogous Hawking radiation and shadows}
\author{Hong Guo}
\email{gh710105@gmail.com}
\affiliation{School of Aeronautics and Astronautics, Shanghai Jiao Tong University, Shanghai 200240, China}
\affiliation{School of Physics and Astronomy,
Shanghai Jiao Tong University, Shanghai 200240, China}

\author{Hang Liu}
\email{hangliu@sjtu.edu.cn}
\affiliation{School of Aeronautics and Astronautics, Shanghai Jiao Tong University, Shanghai 200240, China}
\affiliation{School of Physics and Astronomy,
Shanghai Jiao Tong University, Shanghai 200240, China}
\author{Xiao-Mei Kuang}
\email{xmeikuang@yzu.edu.cn}
\affiliation{Center for Gravitation and Cosmology, College of Physical Science and Technology, Yangzhou University, Yangzhou 225009, China}
\author{Bin Wang}
\email{wang$_$b@sjtu.edu.cn}
\affiliation{School of Aeronautics and Astronautics, Shanghai Jiao Tong University, Shanghai 200240, China}
\affiliation{Center for Gravitation and Cosmology, College of Physical Science and Technology, Yangzhou University, Yangzhou 225009, China}

\begin{abstract}
\vspace*{0.6cm}
\baselineskip=0.5 cm
Various properties of acoustic black holes constructed in  Minkowski spacetime have been widely studied in the past decades. Recently the acoustic black holes in general spacetime  were proposed .  In this paper, we first investigate the basic characteristics of `curved' acoustic black hole in Schwarzschild spacetime, including the quasi-normal modes, grey-body factor and analogous Hawking radiation. We find that the signal of quasi-normal mode is weaker than that of Schwarzschild black hole. Moreover, as the tuning parameter increases, both the positive real part and negative imaginal part of the quasi-normal frequency approach to the horizonal axis, but they will not change sign. This means that all the perturbations could die off
and the system is stable under those perturbations. Since the larger tuning parameter suppresses the effective potential barrier, so it enhances the grey-body factor. The energy emission rate of Hawking radiation does not monotonically increase of the tuning parameter because of the non-monotonicity of the Hawking temperature. Finally, as a first attempt, we study the acoustic black hole shadow. The radius of acoustic shadow becomes larger as the tuning parameter increases, because both the related acoustic horizon and the acoustic sphere become larger. Our studies could help us to further understand the near horizon geometrical features of the black hole. We also expect that our observations could be detected experimentally in the near future.
\end{abstract}

\maketitle
\tableofcontents

\section{Introduction}

The black hole is one of the most intriguing celestial objects in our universe. It plays a significant role in the study of general relativity(GR), thermodynamics, statistics and quantum mechanics. Via the astrophysical detections, it is still difficult to touch the signal of Hawking radiation or anything else with the interaction of quantum field in the gravitational spacetime. The situation turns around when Unruh proposed the acoustic black hole \cite{Unruh1981}, which provides potential connections between astrophysical phenomena and the tabletop experiments.

In the acoustic model of gravity, the equation of motion describes the propagation of sound modes.  The acoustic black hole is formed by a moving fluid with speed exceeding the local sound velocity through a spherical surface. The acoustic horizon is the boundary where the speed of flow equals the local speed of sound. The features including horizon, ergosphere and Hawking radiation of the analogue black holes were explored  in \cite{Visser:1997ux}, inspired by which more efforts on the analogue Hawking radiation were made in \cite{Zhang2011,Vieira:2014rva}. Moreover, the stability of the static or rotating acoustic black holes has been analyzed via computing the quasi-normal modes \cite{Cardoso:2004fi,Nakano:2004ha,Berti:2004ju,Chen:2006zy}. Readers can refer to \cite{Barcelo:2005fc} for a nice review paper about the analogue black holes.

Experimentally, the first realization of a sonic black hole as Bose-Einstein condensate has been reported in \cite{Lahav:2009wx}. More recently, the remarkable experiments \cite{deNova:2018rld,Isoard:2019buh} reported that the thermal Hawking radiation and the corresponding temperature in an analogue black hole were observed. Besides, progress on stimulated Hawking radiation has also been made in  an optical system\cite{Steinhauer2014,Drori2019,Rosenberg:2020jde} and some other mechanics\cite{Guo:2019tmr,Bera:2020doh,Blencowe:2020ygo}.

Thanks to those significant realizations of astrophysical phenomena in the laboratory, acoustic black hole nowadays attracts more and more attentions. More recent extension on the analogue Hawking radiation  can be seen in \cite{Anacleto:2019rfn,Balbinot:2019mei,Eskin:2019tin,Eskin:2019mqi}.
The thermodynamic-like description of the two-dimensional acoustic black hole has been discussed in \cite{Zhang:2016pqx}. The particle dynamics in the acoustic spacetime was also addressed in \cite{Wang:2019zqw}.

Most of the aforementioned studies were based on the acoustic models constructed in the real Minkowski spacetime. Nevertheless, the authors of \cite{Ge:2010wx,Ge:2010eu,Anacleto:2010cr,Anacleto:2011bv,Ge:2015uaa,Ge:2019our} derived the acoustic black holes from the relativistic fluids with the starting of the Abelian Higgs model.  Especially, by fixing curved spacetime geometry, the authors of \cite{Ge:2019our} studied analogue gravity models by considering the relativistic Gross-Pitaevskii (GP)
theory and Yang-Mills (YM) theory. They constructed the acoustic black hole in general curved spacetime. This is significant and interesting because the black holes in our universe could be in the bath of some kind of superfluid or just the cosmological microwave. Moreover, it was addressed in \cite{Sun:2017eph} that the acoustic black hole could also emerge from black-D3 brane based on holographic approach.

In this paper, we are interested in the acoustic black hole in four dimensional Schwarzschild background, which could be one of  the simplest analogue black hole in curved spacetime. We accept that the characteristics appearing in astrophysical black holes, should also appear in their analogous models.
Here we shall concentrate on the basic characteristics near the acoustic horizon, which are
related with the possible observable quantities of the `curved' acoustic black hole.

The first characteristic we shall explore is the frequency of quasi-normal modes(QNM), which governs the relaxation of the sound wave perturbation. The real part of the QNM frequencies describes the oscillations of the perturbation, while the imaginary part indicates the (un)damped of the mode (see \cite{Cardoso:2003pj} and therein for review). The QNMs of astrophysical black holes have been widely studied because it is one of the fingerprints of a gravity theory or other possible deviations beyond GR.
Thus, the study of QNMs could help to test the (in)stability  and further provide the stable regime of parameters in acoustic black hole .

The second characteristic we shall investigate is the grey-body factors, which is  equal to the transmission probability of an outgoing
wave radiated from the black hole event horizon to the asymptotic region\cite{Myung:2003cn,Harmark:2007jy}.  The frequency dependent grey-body factors measures the modification
of the pure black body spectrum. It gives us
significant information about the near-horizon structure of
black holes\cite{Kanti:2002nr}.
Moreover, based on the grey-body factors, we shall further evaluate the energy emission rate of analogous Hawking radiation.

The last characteristic we consider is the acoustic black hole shadow.
Black hole shadow is another fingerprint of the geometry around the black hole horizon. It describes the black hole properties which depend on the gravitational lensing of the nearby radiation \cite{Cunha:2018acu}. Black hole shadow is
important to determine the near horizon geometry and  its properties are widely studied, see for example \cite{Amir:2018szm,Amir:2018pcu,Jusufi:2019nrn,Haroon:2019new,Vagnozzi:2020quf,Allahyari:2019jqz,Khodadi:2020jij,
Konoplya:2019sns,Dokuchaev:2019jqq,Javed:2019rrg,Long:2019nox,Kumar:2018ple,Psaltis:2018xkc,Konoplya:2019xmn} and therein. Moreover, in the experimental side, the Event Horizon Telescope group detected the black hole images with the use of the shadow properties\cite{Akiyama:2019cqa,Akiyama:2019fyp,Akiyama:2019eap}. Moveover, the detection of gravitational waves \cite{Abbott:2016blz} from black holes (or other compact objects) and other observations strongly motivate us to
disclose more near horizon geometry of black holes.
Thus, as a first attempt we will study the acoustic shadow of the `curved' acoustic black hole.
Theoretically, the acoustic shadow is a region of the listener's sky that is left dumb, if there are
sonic sources distributed everywhere but not between the listener and the acoustic black hole.
We expect that this observation could be detected in the analogue black hole experiment.

The structure of this paper is listed as follows. In section \ref{sec=BG}, we review the acoustic black hole in the Schwarzschild spacetime and then present the covariant scalar field equation in this background. In section \ref{sec=QNM}, we compute  the frequencies of  quasi-normal modes and analyze the stability of the sector under scalar field perturbation. Then in section \ref{sec=GBfactor} and section \ref{sec=shadow}, we study the grey-body factor, Hawking radiation and the shadow properties of the acoustic black hole, respectively. The last section is our conclusion and discussion.

\section{Background and the covariant scalar equation}\label{sec=BG}

\subsection{Acoustic black hole in Schwarzschild spacetime }
In this subsection, we shall first briefly review how the curved acoustic black hole emerge from the GP theory \cite{GP-theory}, and then we focus on the Schwarzschild acoustic black hole. For more details, readers can refer to \cite{Ge:2019our} where the acoustic black hole in the general curved spacetime has been constructed.  The action in GP theory is
\begin{equation}
	S=\int d^4x\sqrt{-g}(|\partial_\mu \varphi|^2+m^2 |\varphi|^2-\frac{b}{2}|\varphi|^4),
\end{equation}
where $\varphi$ is a complex scalar field as order parameter; $b$ is a constant and $m^2$ is a temperature dependent parameter assumed as $m^2\sim (T-T_c)$\cite{GP-theory}. The equation of motion for $\varphi$ is reduced as
\begin{equation}\label{kgequ}
\Box\varphi+m^2\varphi-b|\varphi|^2\varphi=0.
\end{equation}
One could fix a static background spacetime
\begin{equation}\label{bgmetric}
	ds_{bg}^2=g_{tt}dt^2+g_{rr}dr^2+g_{\vartheta\vartheta}d\vartheta^2+g_{\phi\phi}d\phi^2,
\end{equation}
and set the scalar field as $\varphi=\sqrt{\rho(\vec{x},t)}e^{i\theta(\vec{x},t)}$. In the fixed spacetime, one could assume the background solution of the scalar field as $(\rho_0,\theta_0)$, then consider the fluctuations around $(\rho_0,\theta_0)$ as
 \begin{equation}\label{coplexPertur}
 \rho=\rho_0+\rho_1~~~\mathrm{and} ~~~\theta=\theta_0+\theta_1.
 \end{equation}
By substituting \eqref{bgmetric}-\eqref{coplexPertur} into
the Klein-Gordon equation \eqref{kgequ} and considering the long-wavelength limit, one can extract two equations. One is the leading order for the background scalar field
\begin{equation}\label{eqbrho}
b\rho_0=m^2-g_{\mu\nu}\partial_\mu\theta_0\partial_\nu\theta_0=m^2-v_{\mu}v^{\mu}
\end{equation}
where in the second equality we have defined $v_0=-\partial_t \theta_0$, $v_i=\partial_i \theta_0$ ($i=r,\vartheta, \phi$). The other  is a relativistic equation governing the propagation of the phase fluctuation
\begin{equation}
	\frac{1}{\sqrt{-\mathcal{G}}}\partial_\mu(\sqrt{-\mathcal{G}}\mathcal{G}^{\mu\nu}\partial_\nu\theta_1)=0.
\end{equation}
From the above fluctuation equation, one can extract and derive  the effective metric $\mathcal{G}_{\mu\nu}$  as
\begin{equation}\label{element}
	\mathcal{G}_{\mu\nu}=\frac{c_s}{\sqrt{c^2_s-v_{\mu}v^{\mu}}}
\begin{pmatrix}
g_{tt}(c^2_s- v_i v^i) & \vdots & {-v_iv_t}\cr
\cdots\cdots\cdots\cdots & \cdot &\cdots\cdots\cdots\cdots\cdots\cdots\cr
 -v_iv_t & \vdots & {g_{ii}}(c^2_s-v_\mu v^\mu)\delta^{ij}+v_i v_j\cr
\end{pmatrix}
\end{equation}
with $c^2_s\equiv\frac{b\rho_0}{2}$.
It is obvious that the metric $\mathcal{G}_{\mu\nu}$ encodes both the information of the background spacetime $ds_{bg}$ and the background four velocity of the fluid $v_\mu$.

Following \cite{Ge:2019our}, we consider $v_t\neq0,v_r\neq0,v_a=0(a=\vartheta,\phi), g_{tt}g_{rr}=-1$ and the coordinate transformation $dt\to dt-\frac{v_tv_r}{g_{tt}(c_s^2-v_rv^r)}dr$. Then the line element of a static acoustic black holes in the background spacetime metric can be reformed from \eqref{element} as
\begin{eqnarray}\label{acousticMetrc1}
	ds^2&=& c_s\sqrt{c^2_s-v_{\mu}v^{\mu}}\bigg[\frac{c^2_s-v_rv^r}{c^2_s-v_{\mu}v^{\mu}}g_{tt}
dt^2+\frac{c^2_s}{c^2_s-v_rv^r}g_{rr}dr^2+g_{\vartheta\vartheta}d\vartheta^2+g_{\phi\phi}d\phi^2\bigg].
\end{eqnarray}
 %\blue{So the metric can be reformed by some transformation as}
%\begin{eqnarray}\label{acoustic}
%ds^2=c_s^2\bigg[(1-v_rv^r)g_{tt}dt^2+\frac{1}{1-v_rv^r}g_{rr}dr^2+g_{\vartheta\vartheta}d\vartheta^2+g_{\phi\phi}d\phi^2\bigg]
%\end{eqnarray}
We shall focus on the Schwarzschild background  spacetime
\begin{eqnarray}\label{schw}
ds^2_{bg}&=&g_{tt}dt^2+g_{rr}dr^2+g_{\vartheta\vartheta}d\vartheta^2+g_{\phi\phi}d\phi^2\nonumber\\
&=&-f(r)dt^2+\frac{dr^2}{f(r)}+r^2(d\vartheta^2+sin^2\vartheta d\phi^2),
\end{eqnarray}
where $f(r)=1-\frac{2M}{r}$. Subsequently, one can consider an orbit of a vortex that falls freely along the radial from infinity
starting from rest outside a Schwarzschild black hole. Then the radial component
$v_r$ is treated as the escape velocity of an observer who maintains a stationary position
at Schwarzschild coordinate radius $r$. It can be set as $v_r\sim \sqrt{2M\xi/r}$ in which $\xi>0$ is required to guarantee the velocity is real. Note that recalling $c^2_s=\frac{b\rho_0}{2}$ and rescaling $m^2 \rightarrow \frac{m^2}{2c^2_s}$ as well as $v^\mu v_\mu \rightarrow \frac{v^\mu v_\mu}{2c^2_s}$, equation \eqref{eqbrho} could give us the relation $v_{\mu}v^{\mu}=m^2-1$. As addressed in \cite{Ge:2019our}, one can work at the critical temperature of GP theory such that  $m^2$ vanishes, and then one has $v_{\mu}v^{\mu}=-1$. Note that to fulfill the relation $v_{\mu}v^{\mu}=-1$, the time component of the velocity in this case can be worked out as $v_t=\sqrt{f(r)+\frac{2M\xi}{r}f(r)^2}$.

To proceed, we rescale $v^\mu v_\mu \rightarrow \frac{v^\mu v_\mu}{2c^2_s}$ in (\ref{acousticMetrc1}). Then by  substituting the metric functions of (\ref{schw}), $v_r\sim \sqrt{2M\xi/r}$ and $v_{\mu}v^{\mu}=-1$ into it, we can rewrite the line element as
\begin{eqnarray}\label{acousticMetrc2}
&&ds^2=\sqrt{3}c^2_s\bigg[-\mathcal{F}(r)dt^2+\frac{dr^2}{\mathcal{F}(r)}+r^2(d\vartheta^2+sin^2\vartheta d\phi^2)\bigg],\\
&&\mathrm{with} ~~~ \mathcal{F}(r)=\bigg(1-\frac{2M}{r}\bigg)\bigg[1- \xi \frac{2 M}{r}\left(1-\frac{2M}{r}\right)\bigg],
\end{eqnarray}
which is the acoustic black hole metric in Schwarzschild background. Here  $\xi>0$ is defined as the tuning parameter and its regime for the existence of acoustic black hole will discuss later soon. It is noticed that \eqref{acousticMetrc2} recovers the Schwarzschild black hole \eqref{schw} as $\xi\to0$, while as $\xi\to+\infty$ the whole spacetime should be an acoustic black hole because the escape velocity $v_r$ goes to infinity. We shall then set $c^2_s=1/\sqrt{3}$ for convenience.
%%%%%%%%%%
\begin{figure}[thbp]
\center{\includegraphics[height=2in,width=3in]{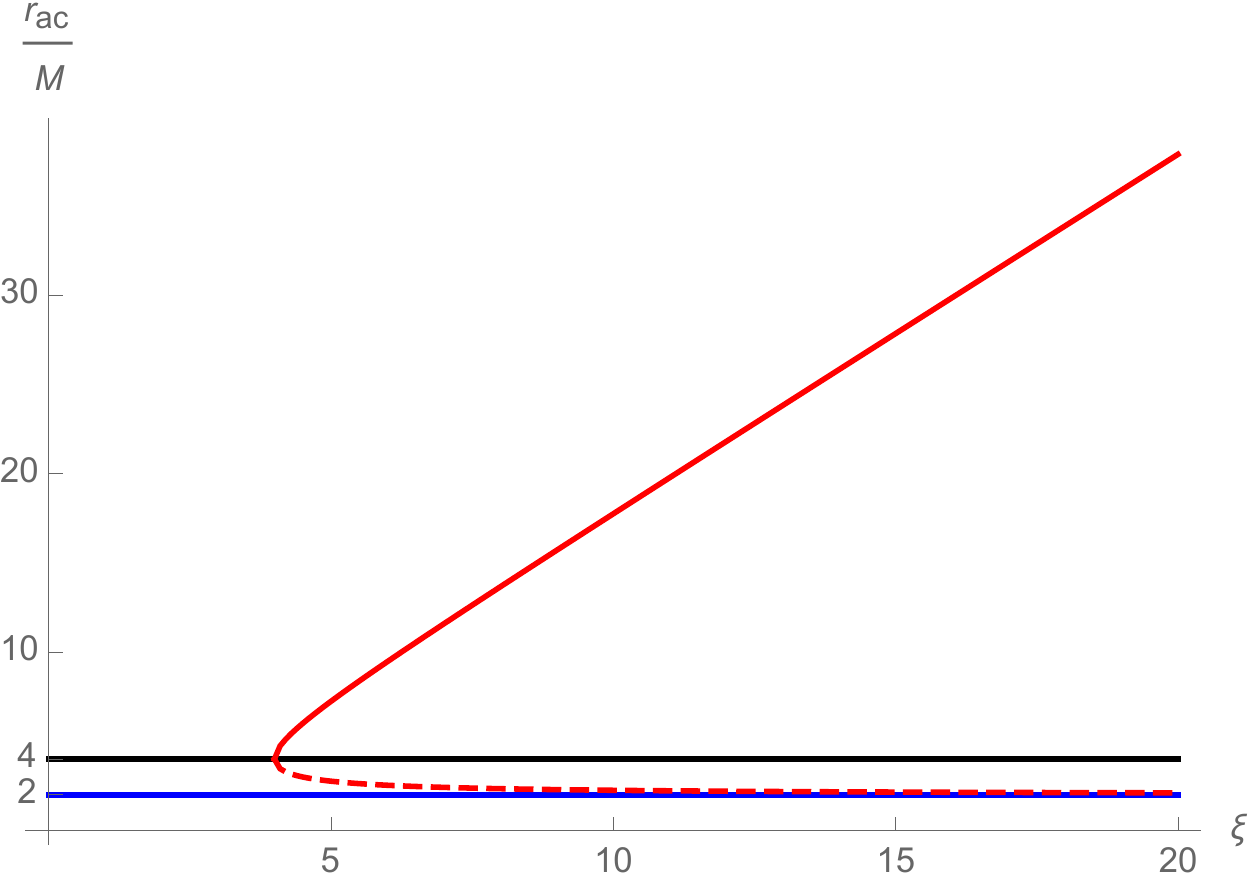}
\caption{The relation of the ratio of $r_{{ac}_{\pm}}/M$ and the tuning parameter $\xi$.}\label{fig-rac}}
\end{figure}

Then to fulfill $\mathcal{F}(r)=0$, we shall obtain three solutions. One is the optical event horizon $r_{bh}=2M$ and the others are  $r_{{ac}_{\pm}}=(\xi\pm\sqrt{\xi^2-4\xi}) M$ for the acoustic black hole.  Recalling $\xi>0$, it is easy to obtain the condition $\xi\geq4$ to make sure the existence of acoustic horizons. When $\xi=4$, the acoustic black hole goes to extreme case with $r_{ac_{-}}=r_{ac_{+}}=4M$. We plot the dependence of $r_{{ac}_{\pm}}$ on $\xi$ in Fig. \ref{fig-rac}. The inner acoustic horizon is confined in  $r_{ac_{-}}\in(2M,4M)$ (the red dashed line), which denotes that the acoustic horizon  locates outside the real black hole as expected. The outer horizon $r_{ac_{+}}$ (the red line) grows monotonously as $\xi$ increases. When $\xi\to\infty$, we have $r_{ac_{+}}\to \infty$ meaning that the sound could not escape from the whole spacetime as we aforementioned.
In a word, in the case with $\xi\geq 4$, an analogue metric would involve in the Schwarzschild spacetime such that the spacetime can be divided into four regions: the inside of black hole is in the regime $r<r_{bh}$; in the regime $r_{bh}<r<r_{ac_-}$ and $r_{ac_-}<r<r_{ac_+}$, the light can escape but the sound cannot; while in the regime $r>r_{ac_+}$, both the light and the sound could escape.  Note that in the following study, the acoustic horizon represents the outer horizon, i.e., we could set $r_{ac}=r_{r_{{ac}_+}}$.

\subsection{Covariant scalar equation }
We are interested in some basic characteristics of this acoustic black holes, including quasi-normal modes and grey-body factors of scalar field as well as the shadow cast. To this end, we consider the minimally coupled massless scalar field as a probe. Its covariant equation is
\begin{equation}
	\frac{1}{\sqrt{-\mathcal{G}}}\partial_\mu(\sqrt{-\mathcal{G}}\mathcal{G}^{\mu\nu}\partial_\nu\psi)=0
\end{equation}
where $\mathcal{G}_{\mu\nu}$ denotes the metric components of the acoustic black holes \eqref{acousticMetrc2}.
Taking the standard ansatz
\begin{equation}
    \psi(t,r,\theta)=\sum_{lm}e^{-i\omega t}\frac{\Psi(r)}{r}Y_{lm}(\theta)
\end{equation}
and introducing the tortoise coordinate $r_*=\int 1/\mathcal{F}dr$, we shall obtain the Schrdinger-like formula
\begin{equation}
	\frac{d^2\Psi}{dr^2_*}+(\omega^2-V(r))\Psi=0
\label{eq1}
\end{equation}
where the effective potential is
\begin{equation}
V(r)= \mathcal{F}\bigg[\frac{l(l+1)}{r^2}+\frac{\mathcal{F}'}{r}\bigg].
\end{equation}
The radial domain of the following study is given by $r\in(r_{ac},\infty)$.

\begin{figure}[thbp]
\center{
\includegraphics[height=2in,width=3.in]{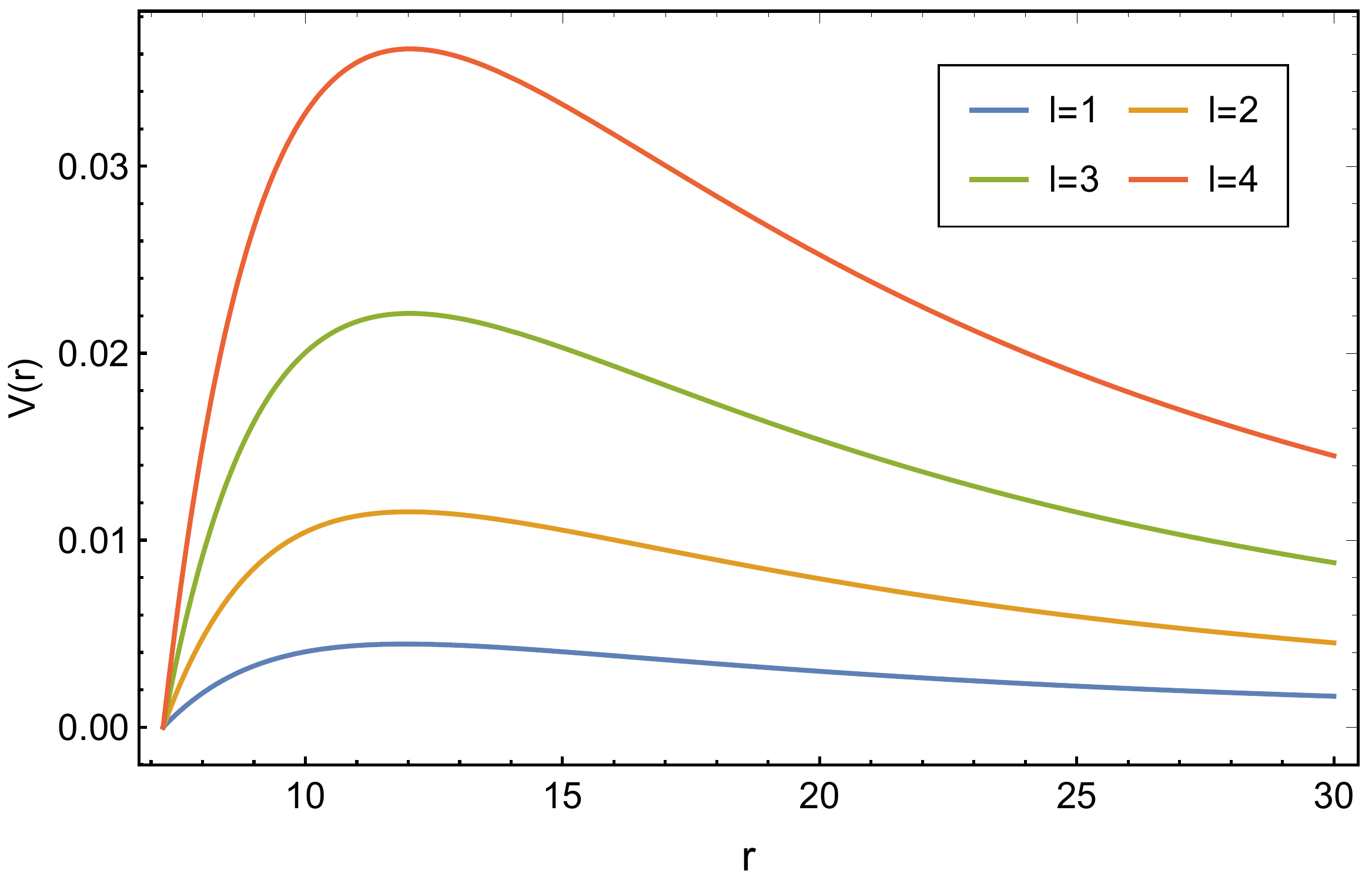}\hspace{0.5cm}
\includegraphics[height=2in,width=3.in]{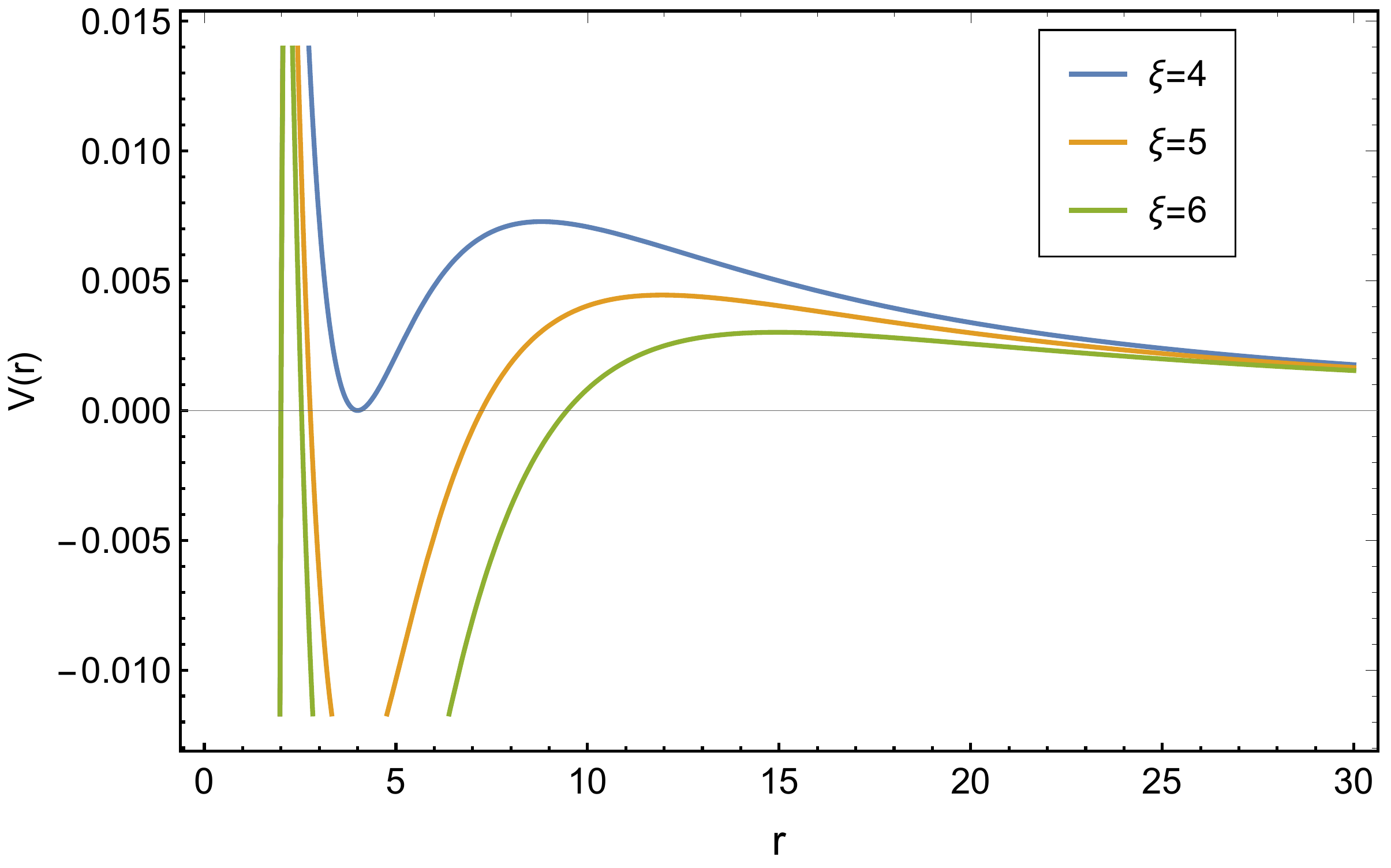}
\caption{The behavior of effective potential with $M=1$. On the left panel we fix $\xi=5$ and take different angular numbers into consideration, while on the right panel we show the potential at fixed $l=1$ for different $\xi$.}\label{potential}}
\end{figure}
The effective potential as a function of $r$ for different cases are present in Fig. \ref{potential}. As the radial coordinate approaches to the near horizon region, the effective potential first shows a barrier and then quickly falls into zero at the acoustic horizon, meanwhile, the tortoise coordinate $r_*$ reaches the infinity $r_* \rightarrow -\infty$.
In the left plot, with fixed $\xi$ and $M$, the position of zero potential is not affected by the angular number, which is reasonable because it has no print on the acoustic horizon. However, the potential barrier is promoted by larger $l$ which is similar to that in Schwarzschild black hole.
On the right plot, as we increase the tuning parameter, the position of zero potential is located at larger radius because the acoustic horizon increases (see Fig. \ref{fig-rac}). In addition, the potential barrier is suppressed  by larger $\xi$. This behavior could be reflected by the near horizon characteristics  as we will show soon.

\section{Quasinormal modes}\label{sec=QNM}
In this section we shall study the quasi-normal modes  from the equation \eqref{eq1}. Thus, we require the purely outgoing waves at infinity and purely incoming waves at the acoustic horizon for the scalar field as
\begin{equation}
	\Psi \sim e^{\pm i\omega r_*}, r_*\rightarrow \pm \infty.
\end{equation}

To calculate the frequency of the QNMs, we employ the semi-analytical WKB method and the asymptotic iteration method (AIM), both of which are widely applied in the study of QNMs. Here we skip the instruction of the two methods,  and the readers can refer to \cite{Konoplya:2019hlu} and  \cite{Cho:2009cj} (and therein) for the details, respectively. It is noticed that even though our results are quite at good convergence in the 6th order WKB approximation, our calculation is accomplished by WKB method upto the 9th order correction for sufficient precision.

The QNM frequencies for small $\xi$ with samples of angular number $l$ and overtone number $n$ are listed in Table \ref{table1} ($l=n=0$), Table \ref{table2}($l=1,n=0$) and Table \ref{table3} ($l=n=1$), respectively. The universal properties of the QNM frequency we can extract from the three tables are:

\begin{itemize}
  \item The real part  $Re(\omega)$ are positive and the imaginal part $Im(\omega)$ are negative, which means the acoustic black hole are stable under the perturbation for small tuning parameters. Moreover, their magnitudes for  the acoustic black hole are quite smaller than that for the Schwarzschild black hole with $\xi=0$. This implies that the signal of the QNM is weaker in acoustic black hole than in the astrophysical  black hole, so the perturbation dies off slower.
  \item With the increasing of  $\xi$, the real part of the frequency decreases. It means that  the oscillation of the scalar field damps. The magnitude of the imaginal part also decreases, denoting the loss of the damping rate. These results imply that the strength of oscillation is damping as the acoustic horizon $r_{ac}$ grows because $r_{ac}$ is larger as $\xi$ increases (see Fig. \ref{fig-rac}). This behavior is reasonable  because the effective potential barrier is suppressed by larger acoustic black hole (see Fig. \ref{potential}).
  \item In each table, the magnitude of $Im(\omega)$ continues decreasing as $\xi$ increases. This indicates that the $Im(\omega)$ may cross zero and changes sign as we continue increasing $\xi$. Since the system is stable only when $Im(\omega)$  is always negative, therefore, to further check the stability, we must study the QNM frequency for general $\xi$.
\end{itemize}
%%%%%%%%%%%%%%%%%%%%%%%%%%%%%%%%%%%%%%%%%5

\begin{table}[!htbp]
\centering
\begin{tabular}{|c|c|c|c|c|c|}\hline
$\xi$& 9th-order WKB &  AIM \\\hline
$0$&$0.11031-0.10496i$&$-$\\\hline
$4$&$0.02836-0.01905i$&$-$\\\hline
$5$&$0.02341-0.01656i$&$0.02351-0.01660i$\\\hline
$6$&$0.01954-0.01471i$&$0.01956-0.01471i$\\\hline
$7$&$0.01668-0.01310i$&$0.01666-0.01307i$\\\hline
$8$&$0.01469-0.01168i$&$0.01449-0.01171i$\\\hline
$9$&$0.01283-0.01066i$&$0.01282-0.01058i$\\\hline
$10$&$0.01148-0.00971i$&$0.01149-0.00964i$\\\hline
\end{tabular}
\caption{ The QNM frequency of acoustic black hole with the mode $l=n=0$.\label{table1}}
\end{table}
%%%%%%%%%%%%%%%%%%%%%%%%%%%%%%%%%%%%%%%%%%
\begin{table}[!htbp]
\centering
\begin{tabular}{|c|c|c|c|c|c|}
\hline
$\xi$& 9th-order WKB   & AIM \\
\hline
$0$&$0.29294-0.09766i$&$-$\\\hline
$4$&$0.08211-0.01744i$&$-$\\\hline
$5$&$0.06391-0.01591i$&$0.06390-0.01591i$\\\hline
$6$&$0.05234-0.01402i$&$0.05234-0.01402i$\\\hline
$7$&$0.04434-0.01240i$&$0.04434-0.01240i$\\\hline
$8$&$0.03848-0.01107i$&$0.03848-0.01107i$\\\hline
$9$&$0.03399-0.00998i$&$0.03399-0.00998i$\\\hline
$10$&$0.03045-0.00908i$&$0.03045-0.00908i$\\
\hline
\end{tabular}
\caption{ The QNM frequency of acoustic black hole with the mode $l=1$ and $n=0$.\label{table2}}
\end{table}
%%%%%%%%%%%%%%%%%%%%%%%%%%%%%%%%%%%%%%%%%%%%%%%%%%%%%%%%%%%%
\begin{table}[!htbp]
\centering
\begin{tabular}{|c|c|c|c|c|c|}
\hline
$\xi$& 9th-order WKB   & AIM \\
\hline
$0$&$0.26431-0.30620i$&$-$\\\hline
$4$&$0.07649-0.05359i$&$-$\\\hline
$5$&$0.06061-0.04869i$&$0.06061-0.04868i$\\\hline
$6$&$0.04947-0.04314i$&$0.04947-0.04313i$\\\hline
$7$&$0.04171-0.03829i$&$0.04171-0.03828i$\\\hline
$8$&$0.03603-0.03426i$&$0.03604-0.03426i$\\\hline
$9$&$0.03171-0.03095i$&$0.03171-0.03095i$\\\hline
$10$&$0.02832-0.02818i$&$0.02832-0.02818i$\\
\hline
\end{tabular}
\caption{ The QNM frequency of acoustic black hole with the mode $l=n=1$.\label{table3}}
\end{table}

\begin{figure}[thbp]
\center{
\includegraphics[height=2in,width=3.in]{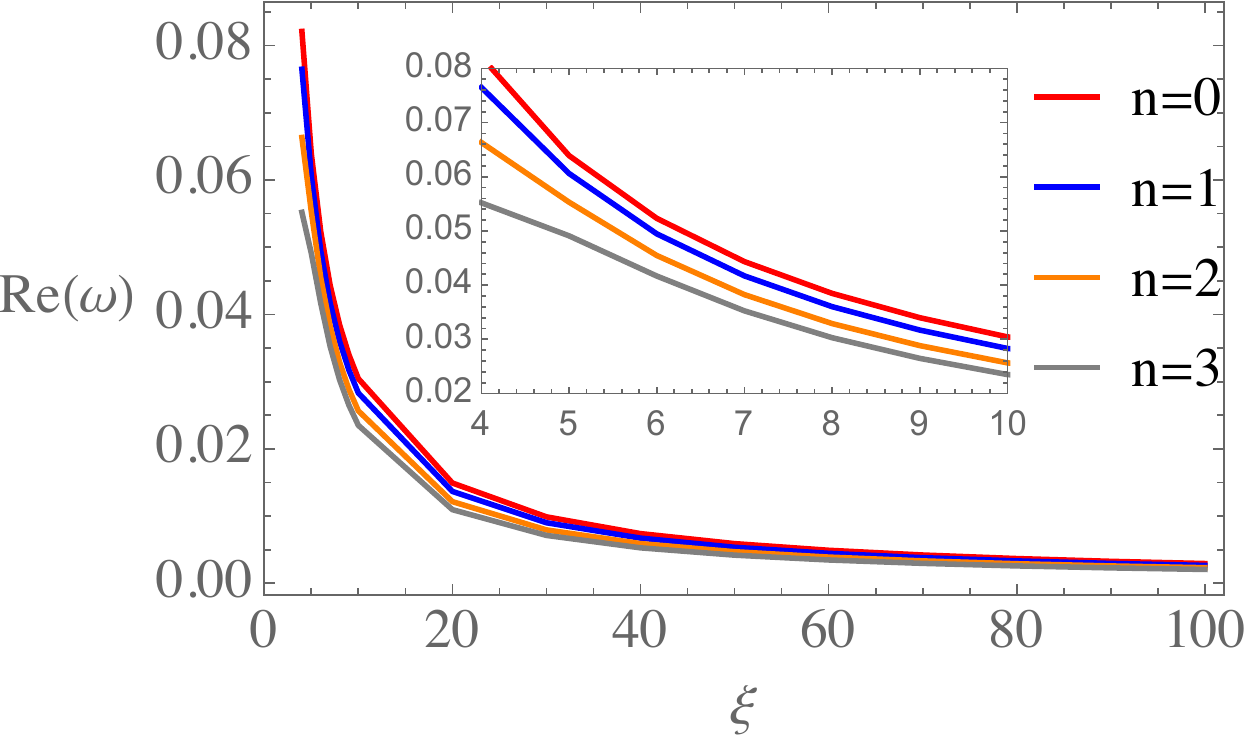}\hspace{0.5cm}
\includegraphics[height=2in,width=3.in]{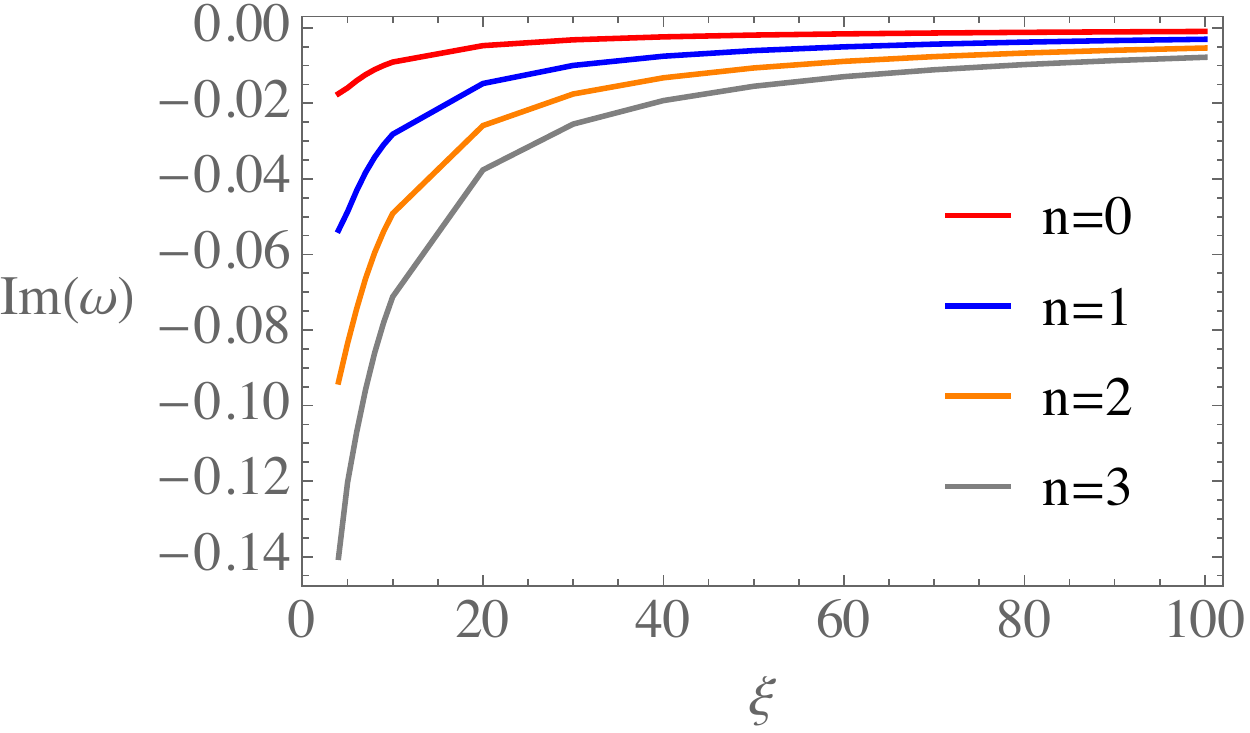}
\caption{The QNM frequency as a function of $\xi$ for different overtone numbers with fixed $l=1$.}\label{frel1}}
\end{figure}

\begin{figure}[thbp]
\center{
\includegraphics[height=2in,width=3.in]{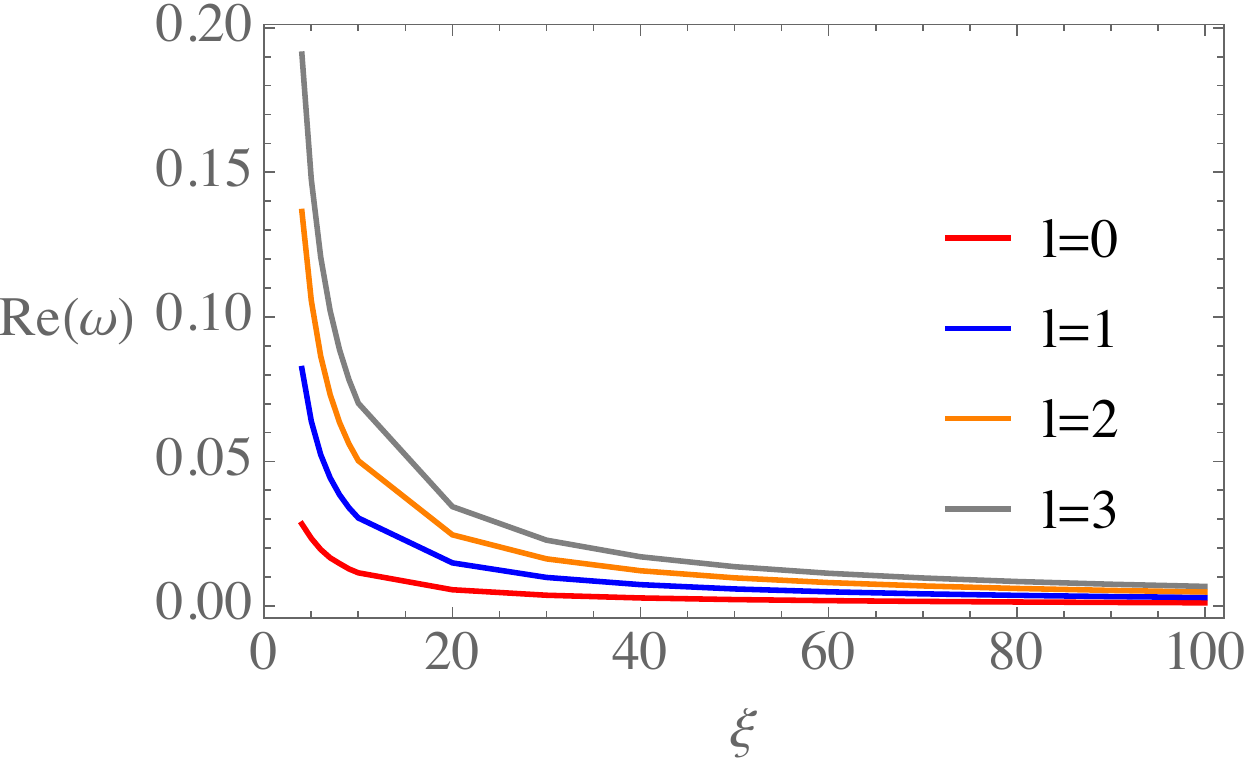}\hspace{0.5cm}
\includegraphics[height=2in,width=3.in]{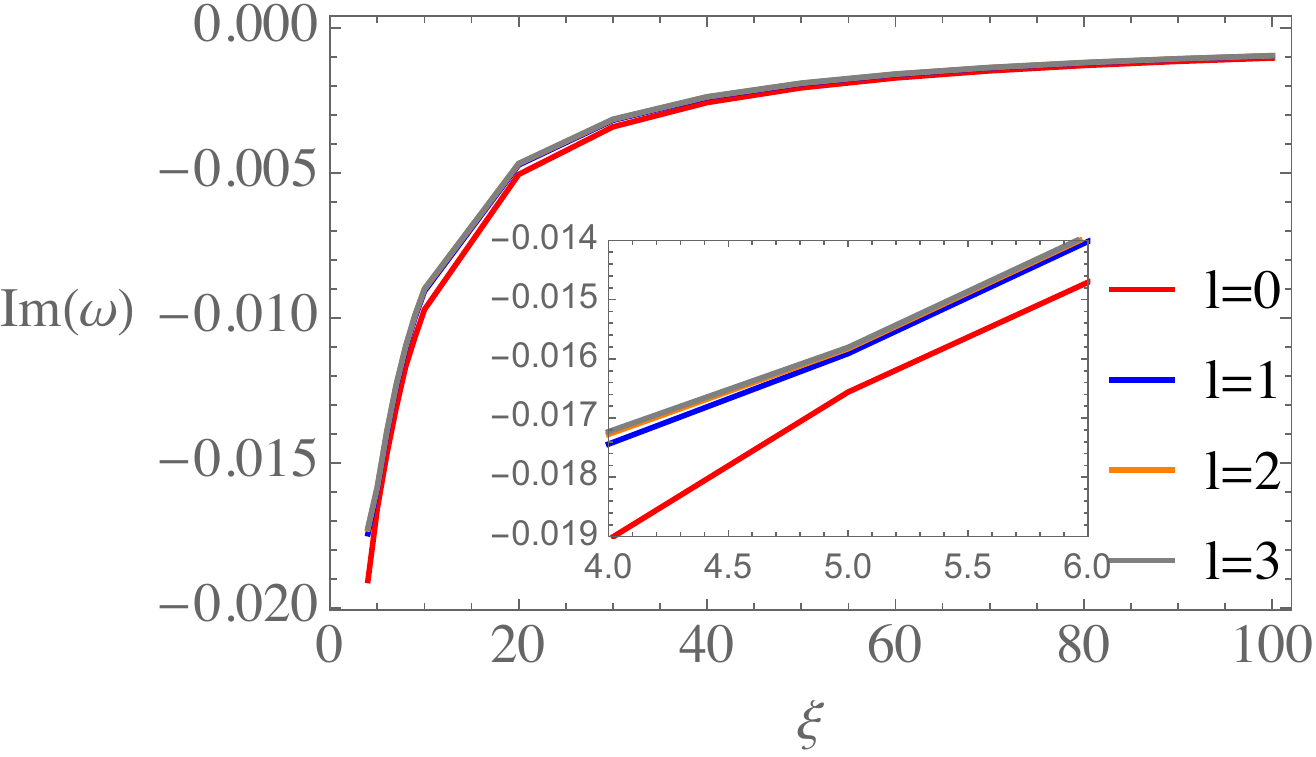}
\caption{The QNM frequency as a function of $\xi$ for different angular numbers with fixed  $n=0$.}\label{fren0}}
\end{figure}

The QNM frequencies as a function of general tuning parameter are shown in Fig. \ref{frel1} and Fig. \ref{fren0} where we choose samples of modes with fixed $l=1$ and $n=0$, respectively. The features of QNM frequency we can obtain from the figures are summarized as:

\begin{itemize}
  \item In both figures, as $\xi$ increases, both the positive real part and negative imaginary part are close to the horizontal axis, but  neither of them changes sign. This indicates that all the perturbation could die off and the acoustic black hole is stable under those perturbations. It is noticed that similar behavior was observed for the counter-rotating waves (negative $l$) as the rotation parameter increases in the rotating acoustic flat black hole \cite{Cardoso:2004fi}.
  \item In Fig. \ref{frel1} with fixed $l=1$, different overtones have different QNM frequency and the difference is more sharp at small $\xi$. Moreover, for larger $n$, both  $Re(\omega)$ and  $Im(\omega)$ are suppressed, which implies that perturbation with larger $n$ die off quicker.  This property is similar as that for acoustic black hole in flat spacetime as well as that for Schwarzschild black hole.
  \item In Fig. \ref{fren0} with fixed overtone, different modes correspond to different QNM frequencies. As $l$ increases, both $Re(\omega)$ and $Im(\omega)$ are enhanced, indicating that the perturbation with smaller $l$ dies off quicker. This property also matches that in acoustic black hole in flat spacetime and Schwarzschild black hole.
\end{itemize}

%%%%%%%%%%%%%%%%%%%%%%%%%%%%%%
%%%%%%%%%
%%%%%%
\begin{figure}[thbp]
\center{
\includegraphics[height=2in,width=3.in]{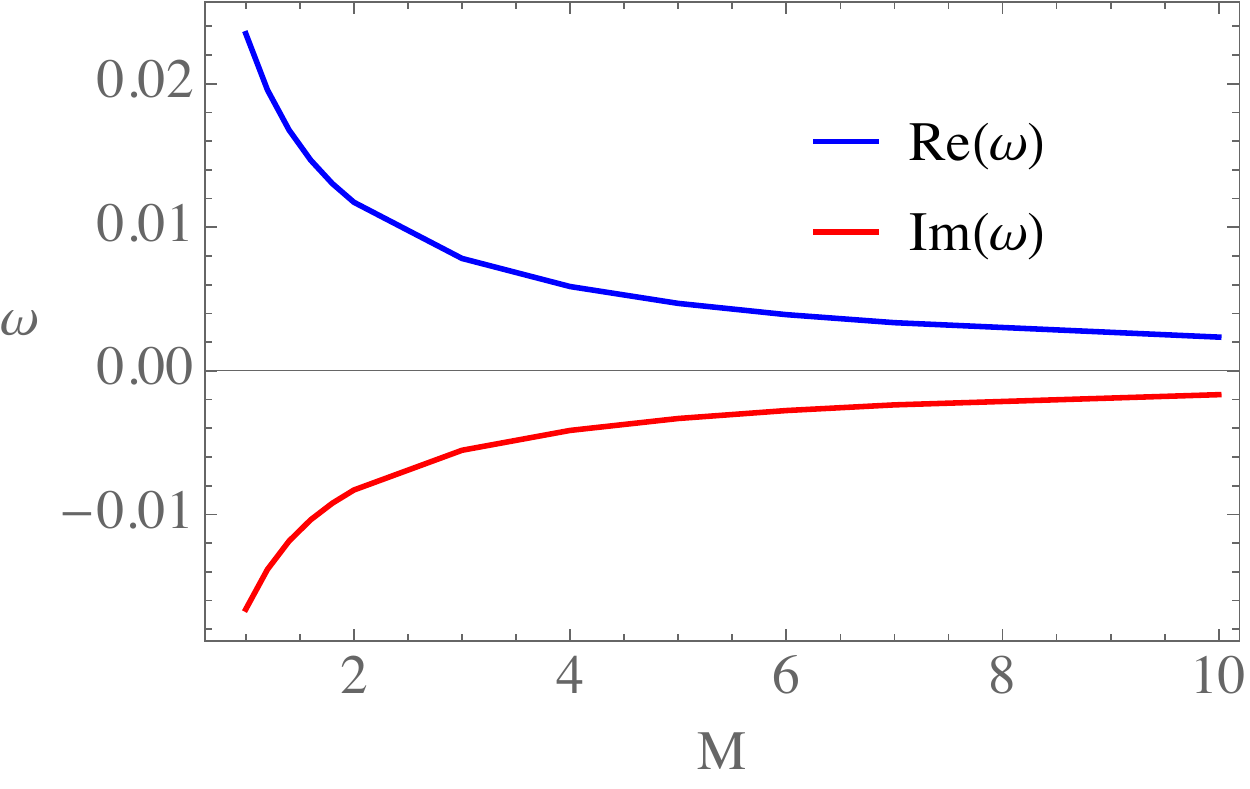}
\caption{The quasi-normal modes as a function of $M$ with fixed $\xi=5$ and $l=n=0$.}\label{freM}}
\end{figure}

Then we further study the effect of mass $M$ on the QNM frequency. The results with fixed $\xi=5$ and $l=n=0$ are shown in Fig. \ref{freM}. It is obvious that as $M$ increases, the positive $Re(\omega)$ decreases and approaches to zero, while the negative $Im(\omega)$ increases and also approaches to the horizonal axis. We did not find the sign changing as we further increase $M$, such that the sector is stable.  This behavior indicates that the existence of heavier black holes or other compact objects would restrain the oscillation amplitude of scalar field, even though it could die off slower. It is noticed that the rule is similar as the effect of $\xi$  because both larger $M$ and $\xi$ corresponds to larger acoustic black hole.
%\red{Does the work \cite{Cardoso:2004fi} consider the effect of M, is the result consistent with ours? What is the effect of M on QNM in Schwarzschild black hole. Note: Work \cite{Cardoso:2004fi} only discuss the acoustic black hole, there is no real black hole parameter $M$; While in the Schwarzschild case, the parameter $M$ is the black hole parameter not the external parameter, there is a linear relationship between QNMs and $M$.}
%
%In the next section, the discussion of the Hawking radiation is accomplished in detail. This part of features can give us more information about the near horizon region.

\section{Grey-body factor and Hawking radiation}\label{sec=GBfactor}
In this section we investigate the grey-body factor and analogue Hawking radiation of the acoustic black hole.
The existence of an acoustic horizon implies the emission of a thermal flux of phonons, named analogue Hawking radiation, and the temperature is proportional to the gradient of the velocity field at the acoustic horizon.

There are plenty of approaches proposed to study the Hawking radiation for astrophysical black hole. It is known that the radiation is not exactly of black-body type since the particles which are created in the vicinity of event horizon without enough energy can  not penetrate the potential barrier. So only part of the particles can be observed at infinity, which makes it just a scattering problem. Thus, to get the transmission probability of particles, we solve the wave equation outside the black hole (acoustic black hole in our present consideration) and calculate the scattering coefficient by which the grey-body factor can be obtained.
We use the grey-body factor to describe the transmission of particles through the potential, and thus work out the energy radiation rate based on the obtained grey-body factor. Note that the effective potential presents a barrier which monotonically decreases as the radius coordinate $r_{*}$ approaches both infinities.  This behavior allows us to use WKB approach to compute the grey-body factor.

We should consider the wave equation Eq.(\ref{eq1}) with the boundary condition allowing the incoming waves from infinity. This is different from that for computing QNMs, where only outgoing waves are allowed in the infinity. The scattering boundary condition is given by
\begin{align}
&\Psi=T e^{-i\omega r_\ast}, \quad \quad \quad r_\ast \rightarrow -\infty,\\
&\Psi=e^{-i\omega r_\ast}+Re^{i\omega r_\ast}, r_\ast \rightarrow +\infty
\end{align}
where  $R$ and $T$ are the reflection and transmission coefficients satisfying  $|T|^2+|R|^2=1$. The grey-body factor is then given by the transmission coefficient for each angular number as\cite{Iyer:1986np}
\begin{equation}
|A_l|^2=1-|R_l|^2=|T_l|^2\quad \mathrm{and}\quad |T_l|^2=(1+e^{2i\pi K})^{-1}
\end{equation}
where $K$ is determined by the equation
\begin{equation}\label{eq-K}
K=i\frac{\omega^2-V_0}{\sqrt{-2V_0''}}-\sum_{i=2}^{i=6}\Lambda_i(K).
\end{equation}
Here $V_0$ and $V_0''$ denotes the maximal value of the effective potential and its second derivative with respective to the tortoise coordinate at the maximum, respectively; and $\Lambda_i$ are the higher WKB corrections which are dependent on $K$ and up to 2$i$th order derivative of the potential at its maximum\cite{Schutz:1985km,Iyer:1986np,Konoplya:2003ii}. We briefly review how to derive $\Lambda_{i}(K)$ in Appendix \ref{app}.

Once the grey-body factor is at hands, we can then study the Hawking radiation by evaluating the energy emission rate which is connected with the grey-body factor via \cite{Hawking:1974sw}
\begin{equation}
\frac{dE}{dt}=\sum_{l}N_l|A_l|^2\frac{\omega}{e^{\omega/T_H}-1}\frac{d\omega}{2\pi}.
\end{equation}
In the above definition, $T_H$ is the analogue Hawking temperature defined as $T_H=-\mathcal{F}'(r_{ac})/4\pi$ and $N_l$ are the multiplicities satisfying  $N_l=2l+1$ for the scalar field.

\begin{figure}[thbp]
\center{
\includegraphics[height=2in,width=3.in]{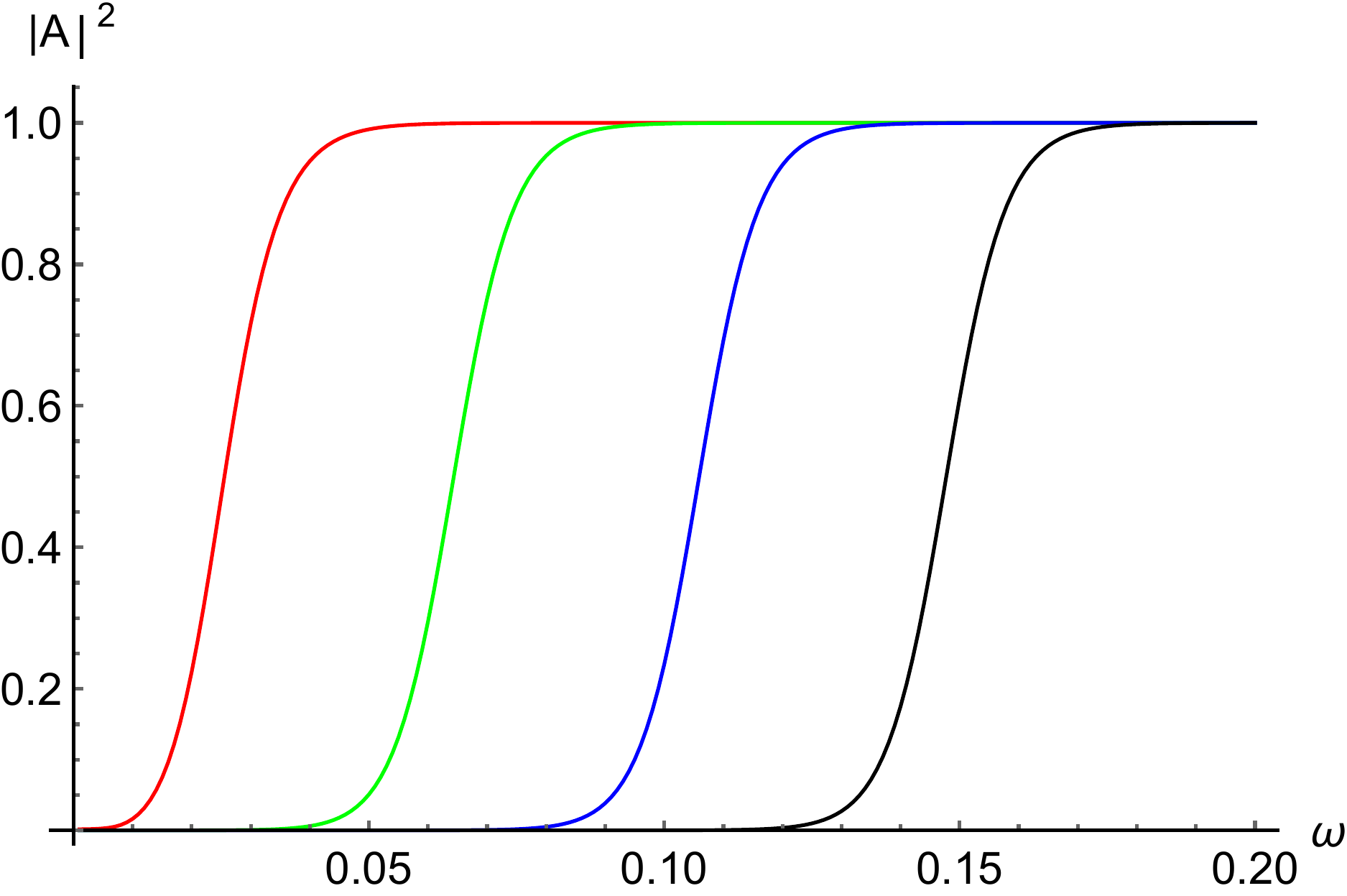}\hspace{0.5cm}
\includegraphics[height=2in,width=3.in]{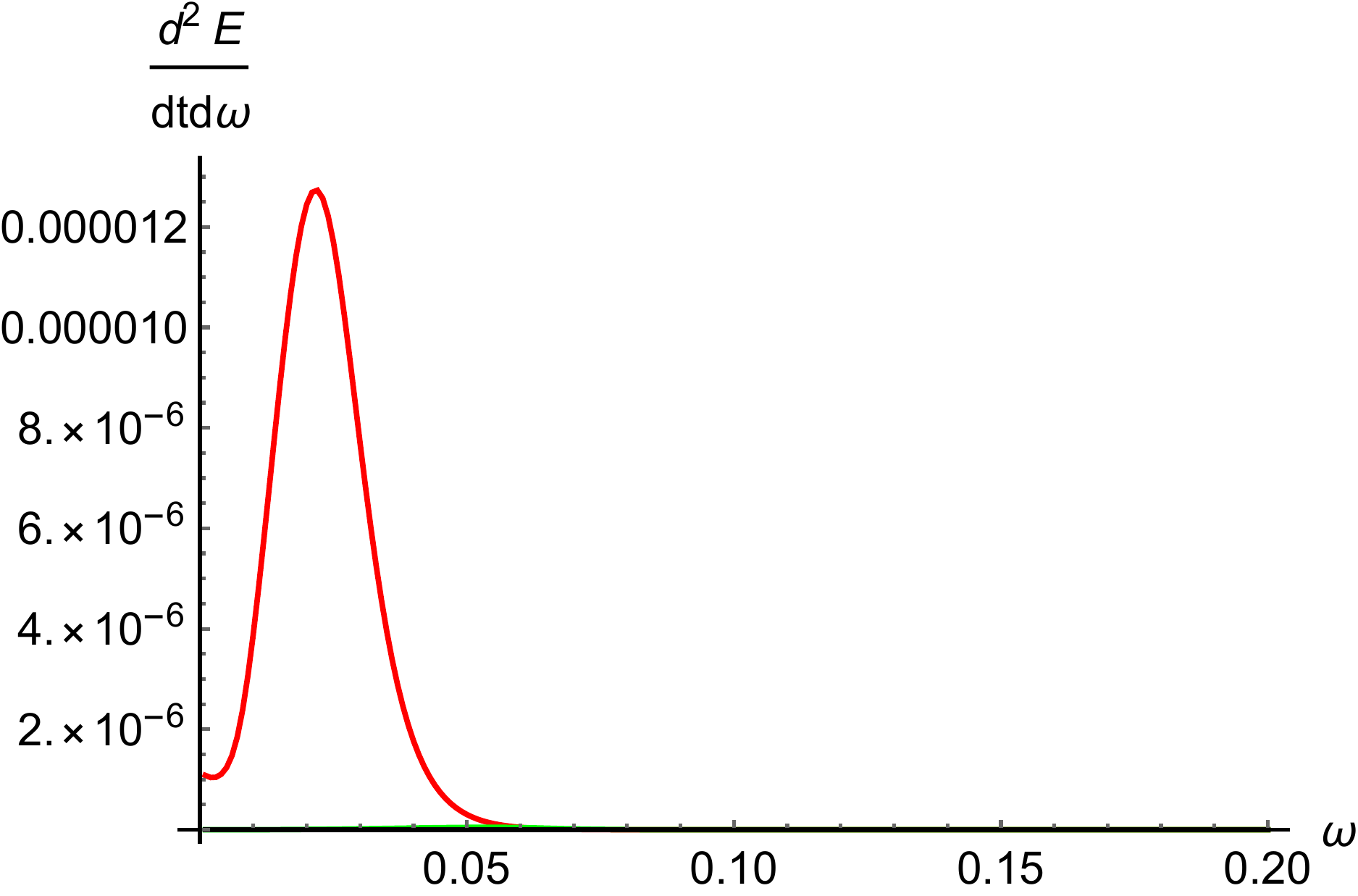}
\caption{The left panel shows the grey-body factor  and the right panel shows the partial energy radiation rate for different angular numbers. For both panels, we fix $\xi=5$, and the red, green, blue and black lines correspond to angular numbers $l=0,1, 2$ and  $3$, respectively.  On the right plot, the radiation rate for $l>0$ are too weak to be observable.}\label{grfactor1}}
\end{figure}

\begin{figure}[thbp]
\center{
\includegraphics[height=2in,width=3.in]{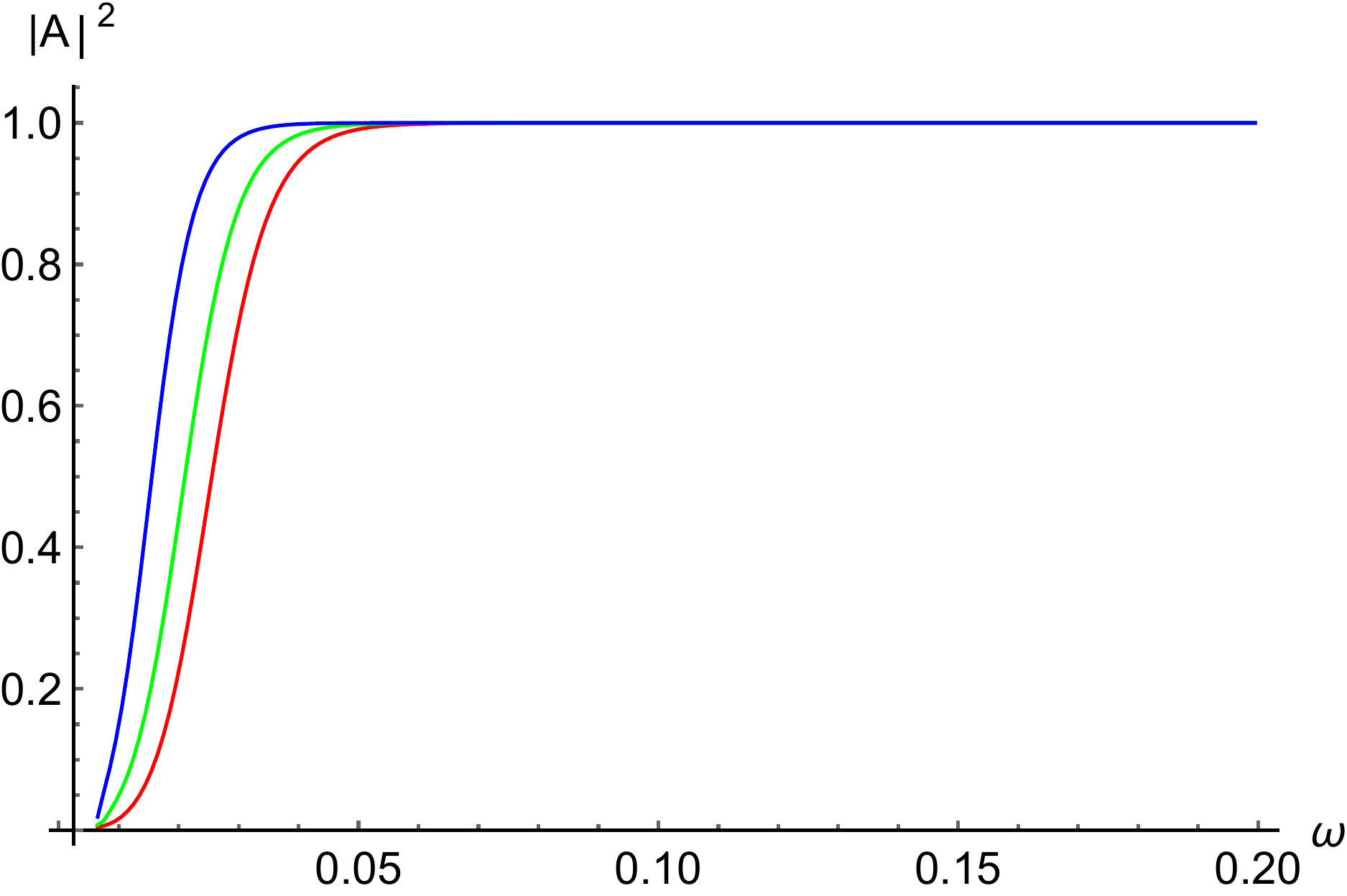}\hspace{0.5cm}
\includegraphics[height=2in,width=3.in]{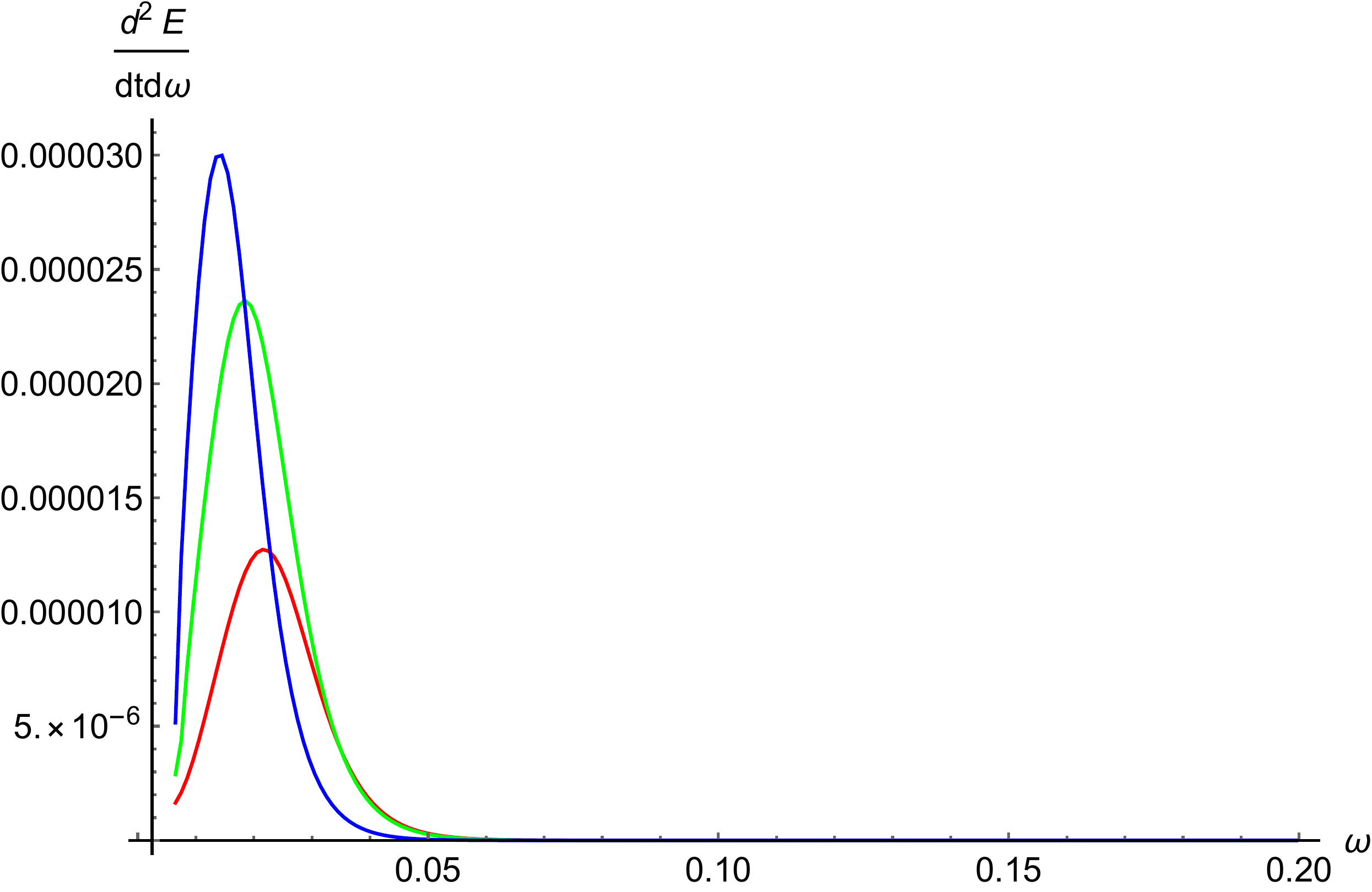}
\caption{The left panel shows the grey-body factor and the right panel shows the partial energy radiation rate  for different $\xi$. For both panels, we fix $l=0$, and the red line, green line and blue line correspond to  $\xi=5$, $\xi=6$ and $\xi=8$, respectively.}\label{grfactor3}}
\end{figure}

\begin{figure}[thbp]
\center{
\includegraphics[height=2in,width=3.in]{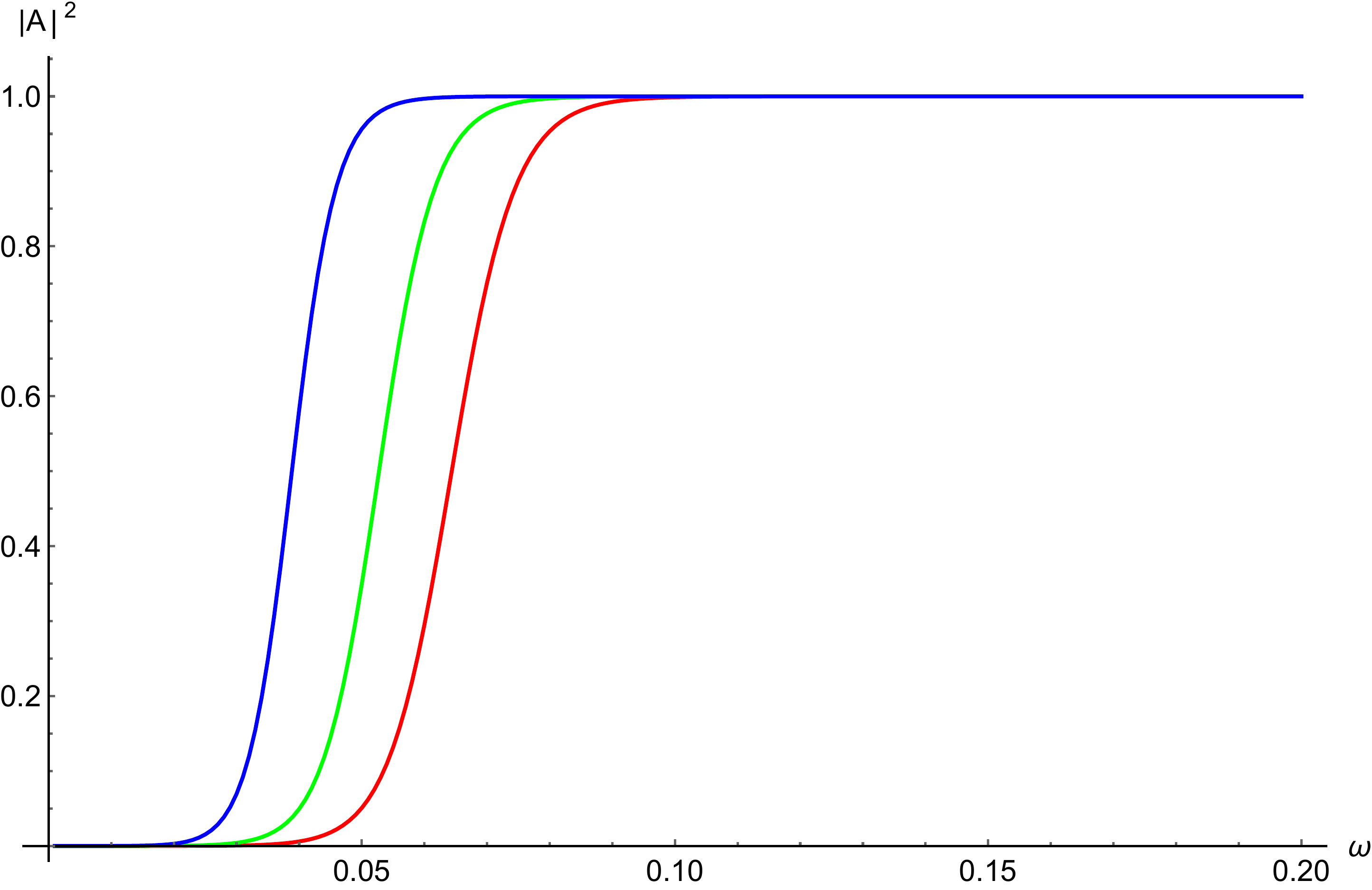}\hspace{0.5cm}
\includegraphics[height=2in,width=3.in]{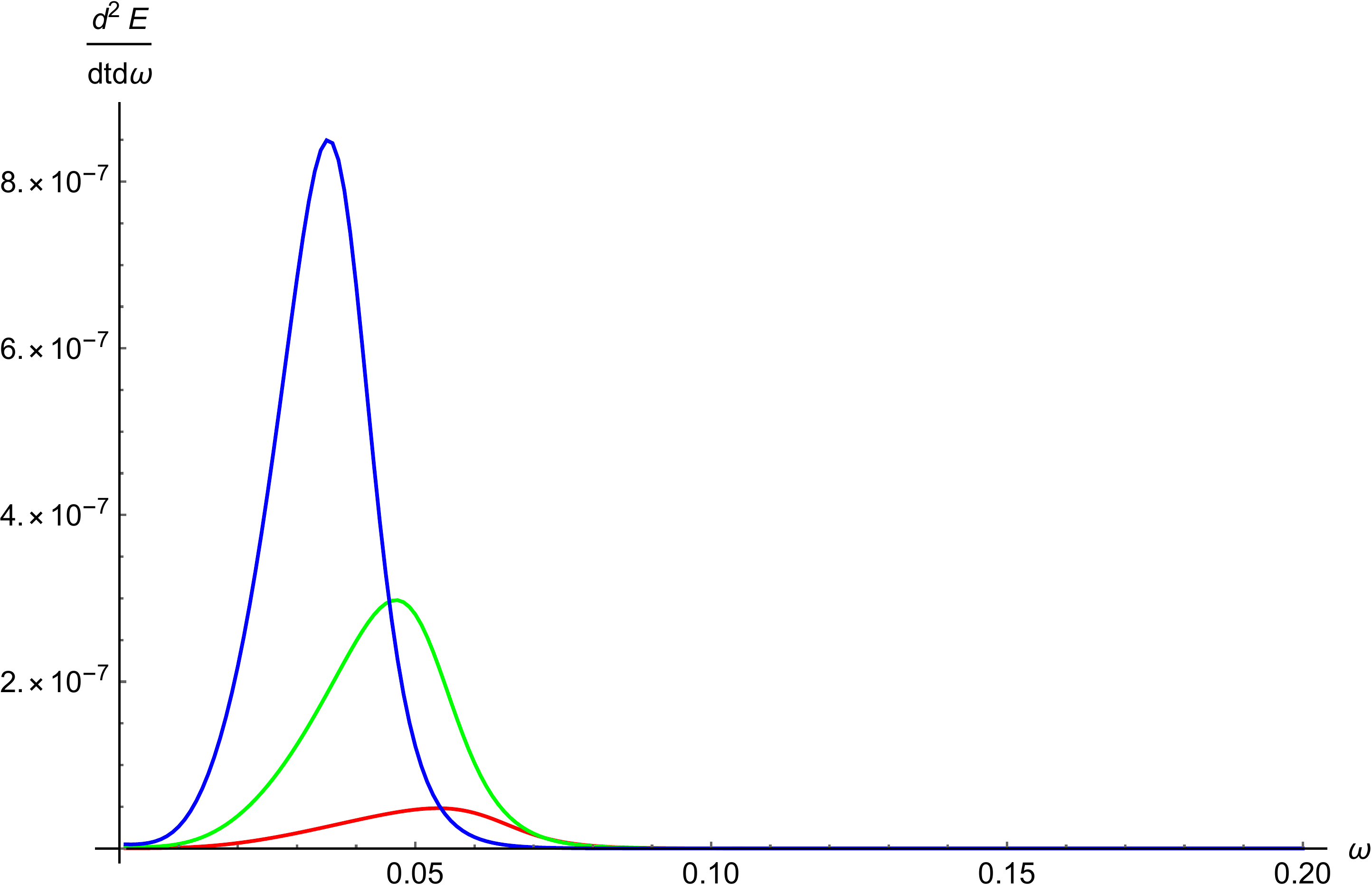}
\caption{The left panel shows the grey-body factor and the right panel shows the partial energy radiation rate  for different $\xi$. For both panels, we fix $l=1$, and the red line, green line and blue line correspond to  $\xi=5$, $\xi=6$ and $\xi=8$, respectively.}\label{grfactor2}}
\end{figure}

Then, we shall employ the 6th order WKB method  to calculate the grey-body factor and energy emission rate of Hawking radiation. It is noticed that this method was employed to study the properties of Hawking radiation in various models, see for examples \cite{Konoplya:2010kv,Volkel:2019ahb,Konoplya:2019hml,Konoplya:2019ppy} and therein. Our results are shown in Fig. \ref{grfactor1}-\ref{grfactor2}.
%%%%%%%%%%%%%%%%%%%%%%

In Fig. \ref{grfactor1} we fix the tuning parameter $\xi=5$ and study the effect of the angular number $l$. The left panel shows that the larger frequency corresponds to the  higher grey-body as a natural consequence of the fact that the particles with larger energy are more likely to penetrate the potential barrier. On the other hand, it is obvious that larger angular number leads to a lower grey-body factor. This result can be intuitively explained by the effective potential which has higher barrier for larger $l$ (see Fig. \ref{potential}), such that the particles are more likely to be reflected by the potential. The energy emission rate of the Hawking radiation is shown in the right panel. It is observed that the mode with $l=0$ dominates the Hawking radiation while  the contribution of modes with higher $l$ is very small and negligible.
%%%%%%%%%%%%%%%%%%%%%%%%%%%%%

%%%%%%%%%%%%%%%%%%%%%%%%%%%%%%%%
\begin{figure}[thbp]
\center{
\includegraphics[height=2in,width=3.in]{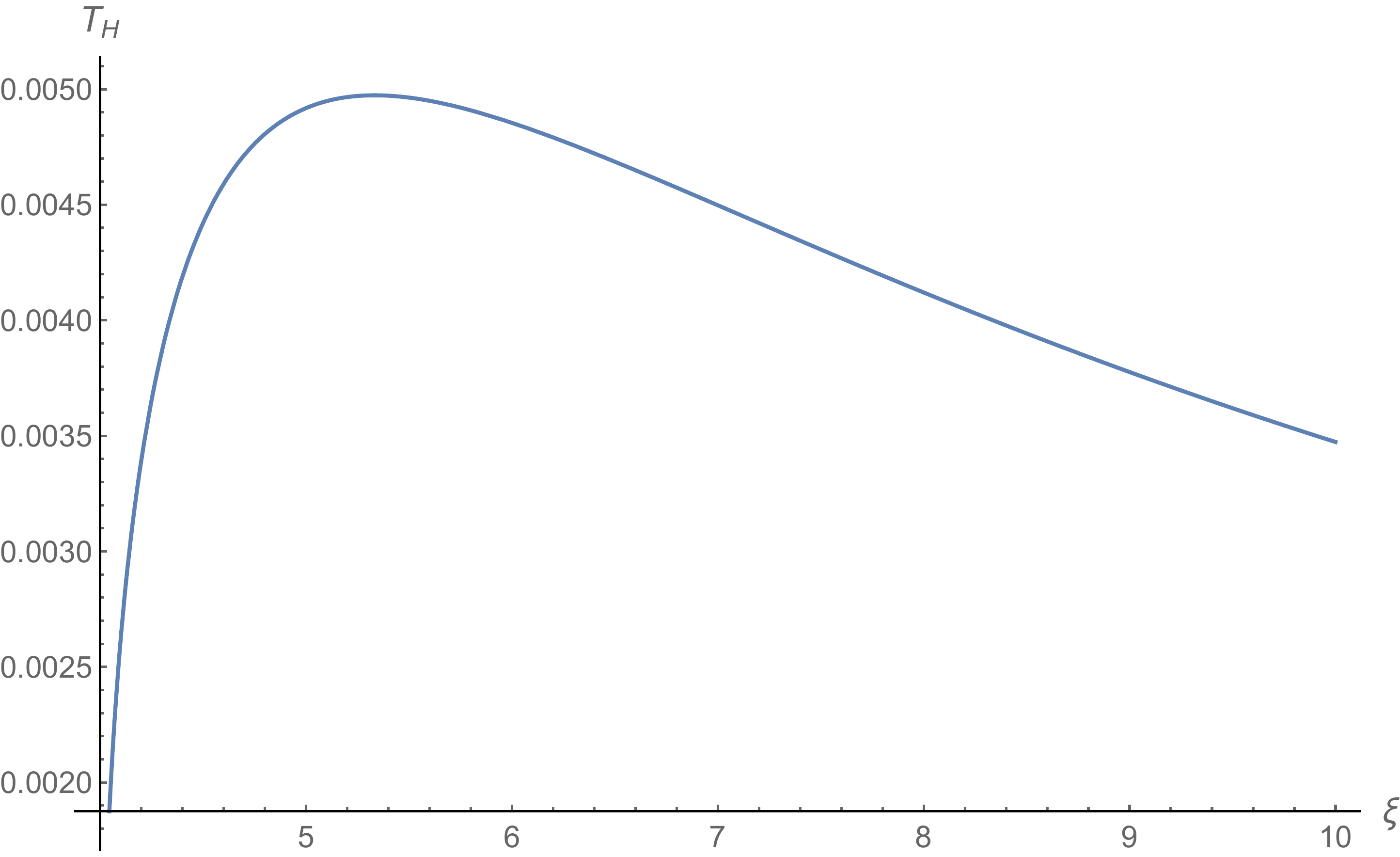}
\caption{The behavior of Hawking temperature as a function of $\xi$. Here we take $M=1$.}\label{temp}}
\end{figure}

In Fig.\ref{grfactor3} and Fig.\ref{grfactor2}, we choose different tuning parameters and fix the angular number $l=0$ and $l=1$, respectively. The behaviors in the two figures are qualitatively similar with an obvious exception that the energy radiation rate for $l=0$ is much stronger than that for $l=1$. The left panels in the figures show that the grey-body factor is enhanced by larger $\xi$ because of the lower potential barrier as we have shown in Fig. \ref{potential}. In the right panels, the emission rate at the low frequency region for larger $\xi$ is larger, but when $\omega$ grows larger, this picture will be changed as a result of its Hawking temperature dependence. It is shown in Fig. \ref{temp} that the Hawking temperature grows from zero which corresponds the extremal black holes at $\xi=4$  to a maximum at $\xi=16/3$, and then it decreases to suppress the energy emission rate.

\section{Acoustic black hole shadow}\label{sec=shadow}
The black hole shadow in GR is one of the optical properties. As a first attempt, we shall study the analogous behavior in the acoustic black hole. We treat it as ``acoustic shadow" which describes the property of the sound waves. Analogous with the gravitational lensing and the related properties of astrophysical black hole, there exists the inmost unstable sound wave orbit and we call it ``acoustic sphere" instead of ``photon sphere". Beyond the acoustic sphere, the sound waves are absorbed by the acoustic black hole, so the acoustic sphere describes the ``audible" boundary of the sound waves near the acoustic black hole horizon region. This property is related with the acoustic shadow.

Moreover, it is known that the shadow for static and spherical symmetric black holes  also has spherical symmetry.
The shape of shadow would be more complex when the rotation of the black hole is involved. Thus, here for the static and spherical symmetric acoustic black hole \eqref{acousticMetrc2}, we simply study the shadow radius and discuss how the tuning parameter and the black hole mass shall affect the shadow radius.

To proceed, we follow the designations of \cite{Perlick:2015vta} and  find the radius of the ``acoustic sphere" $r_{ah}$ by solving the following equation
\begin{equation}
	\frac{dh^2(r)}{dr}=0
\end{equation}
where the function $h(r)$ is defined as  $h(r)=\sqrt{r^2/\mathcal{F}(r)}$. Then for a distant static listener locating at $r_L$, the detected  radius of the acoustic shadow is
\begin{equation}
	r_{sh}=\frac{h(r_{ah})r_L}{h(r_L)}.
\end{equation}

\begin{figure}[thbp]
\center{
\includegraphics[scale=0.6]{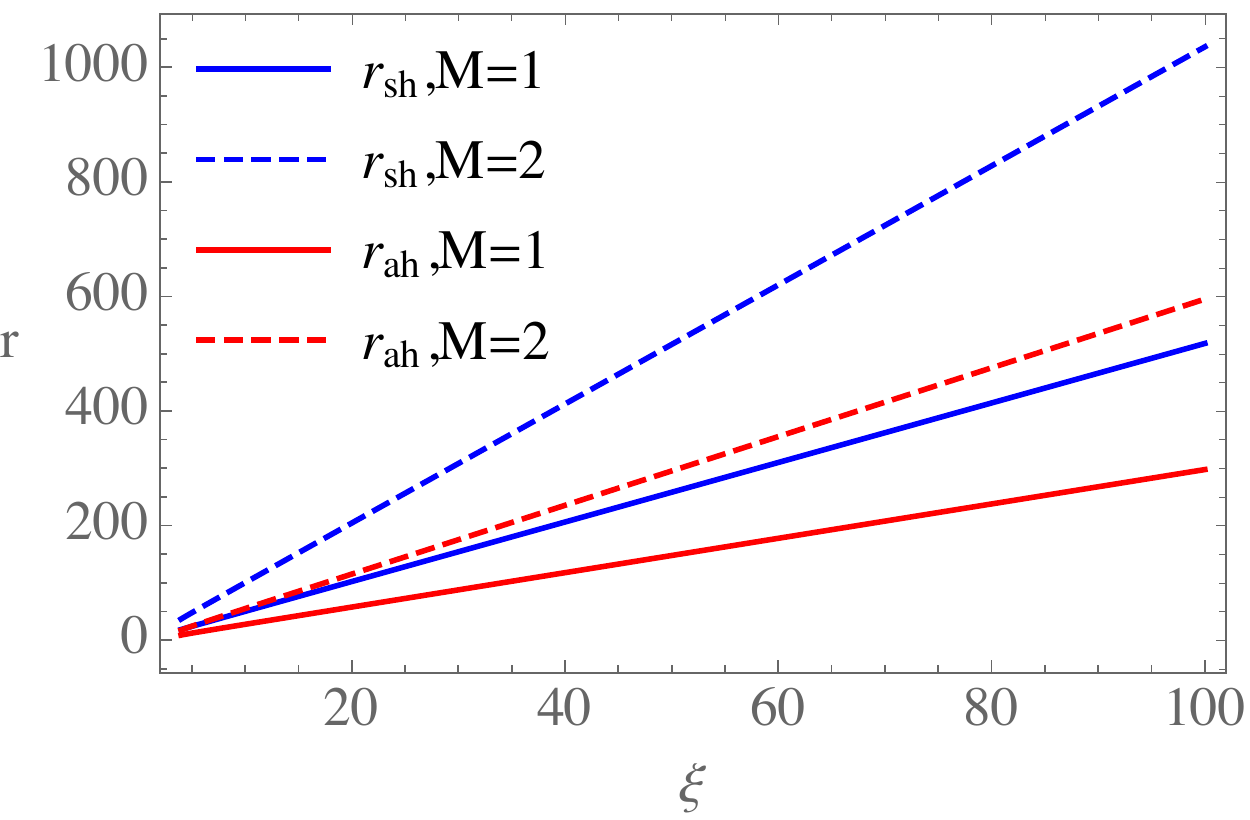}
\caption{The shadow radius $r_{sh}$ and the acoustic sphere $r_{ah}$ changing with the tuning parameter $\xi$.}\label{shadow}}
\end{figure}

To study the properties of the acoustic shadow,  we assume the static listener is far away from the vicinity of acoustic horizon so that we have $\frac{r_L}{h(r_L)}\approx 1$.  The radius of the acoustic sphere  and the acoustic shadow as functions of the tuning parameter are given in Fig. \ref{shadow}. It is obvious that with fixed $M$, both $r_{sh}$ and $r_{ah}$ increase almost linearly as $\xi$, but the slope for the shadow radius is larger than that for the acoustic sphere. Moreover, in the figure, we can also see the influence of the mass parameter $M$ on the radius that the larger $M$ corresponds to both larger shadow radius and acoustic radius. This is reasonable because the increase of $M$ could enlarge the acoustic horizon. It is noticed that the radius of acoustic sphere is much larger than the photon sphere for Schwarzschild black hole which is $3M$.

%\red{Simplify the formula of shadow radius, $r_{sh}$ is given by the acoustic sphere $r_{ah}$ divided by the root of $f(r_{ah})$. In order to analyze the behavior of the shadow radius, we find that the value of $f(r_{ah})$ is approach to a constant $\frac{1}{3}$ in the limit of $\xi \rightarrow \infty$ and that without any relation of $M$. In conclusion, the contribution to the acoustic shadow radius comes from the behavior of the acoustic sphere $r_{ah}$ just shown in Fig.\ref{shadow}.}
%%%%%%%%%%%%%%%%%%%%%%%%%%%%%

\section{Conclusion and discussion}
In this paper, we explored the near-horizon characteristic of the `curved' acoustic black hole in the Schwarzschild spacetime. By considering the minimal coupling massless scalar field, we studied the quasi-normal mode, grey-body factor and analogous Hawking radiation of the sector. Moreover, as a first attempt, we also studied the acoustic shadow which is analogous to the photon shadow caused by the bent light ray in general relativity.

We computed the frequencies of QNMs with the use of WKB method upto ninth order corrections as well as the asymptotic iteration method. Our results showed that the signal of QNMs in this acoustic black hole is weaker than that in Schwarzschild black hole. Moreover, the real part is always positive  while the imaginal part of the QNM frequency is negative, and both of them get closer to the horizonal axis as the tuning parameter increases. The sign of the QNM frequency does not change, which implies that all the perturbations would die off and the acoustic black hole is stable under those perturbations. It would be interesting to further test the stability under other types of perturbations around the acoustic black hole, which will be present elsewhere.

We then investigated the analogous Hawking radiation of the acoustic black hole. We employed the WKB approach to solve the scalar equation as a scattering problem. Both the grey-body factor and the energy emission rate of Hawking radiation are affected by the angular number and the tuning parameter  which correspond to different properties of the potential barrier. Especially, the grey-body factor is enhanced by the larger tuning parameter because it corresponds to lower potential barrier. The energy emission rate of Hawking radiation is not a monotonic function of the tuning parameter due to the non-monotonicity of the Hawking temperature.

Finally, we studied the acoustic shadow of acoustic black hole. Since the acoustic black hole we considered is static and spherical symmetric, so the acoustic shadow also has spherical symmetry. Thus, we simply analyzed the acoustic shadow radius, and we found that the radius of shadow becomes larger as the tuning parameter increases. This is acceptable  because as the parameter increases, both the acoustic horizon and the acoustic sphere increases.

It is worthwhile to mention that our work is the first attempt to study ``black hole shadow" in the acoustic black hole.
Though here we worked with the techniques developed in the optical case, one could expect the experimental interest of the ``acoustic shadow" due to the following two aspects. On one hand, as we mentioned in the introduction, the experimental simulation of the analogous black hole is significant for us to understand astrophysical phenomena. The acoustic shadow is a direct observable quantity which would shed light on the experimental simulation of the acoustic black hole. Especially, the acoustic shadow is a great indicator to describe the near acoustic horizon region in the laboratory, so the study of acoustic shadow could help us further `touch' the essences of the real black hole.
On the other hand, the study on this topic could  be a good chance to detect the motion of the sound waves in a medium.  The experimental study of the acoustic shadow could be a possible clue to understand the similarity  and differences between the sonic fluid and the black hole geometry. Indeed, there has been some attempts in this direction, for instance, the sound wave shadow zone in a stratified ocean was studied in \cite{pekeris1946theory} many yeas ago. Inspired by this work, people found the zero-reflection effects of the sound waves \cite{badulin_shrira_tsimring_1985,badulin_shrira_1993,Mironov} in which the authors associated the properties with the ``black holes" effect.

It is expected that most of the black holes in the galactic center rotate. Moreover, rotating black holes are closely related with the two important directions: the gravitational waves and black hole shadows, which open new
windows for us to understand the universe. Thus,
it would be significant to extend our studies into the acoustic black hole with rotation in the curved spacetime, and then further study the connection between QNM frequency and shadow which was proposed in \cite{Stefanov:2010xz}.
We also expect that our theoretical results could be observed in analogous black hole experiment in the near future. This could help us to further understand the structure of near horizon geometry of astrophysical black holes.

\begin{acknowledgments}
This work is supported by the Natural Science Foundation
of China under grant Nos.11705161 and 11835009, Fok Ying Tung Education Foundation under grant No.171006 and the
Natural Science Foundation of Jiangsu Province under grant No.BK20170481.
\end{acknowledgments}

\appendix

\section{Higher Order Correction Terms $\Lambda_i(K)$ of WKB Approach}\label{app}
In this Appendix, we shall follow \cite{Iyer:1986np} and briefly introduce how to fix the higher order correction terms $\Lambda_i(K)$  as we show in equation \eqref{eq-K} for the WKB method. $\Lambda_2$ and $\Lambda_3$ are derived in the following and one can refer to  Ref. \cite{Konoplya:2003ii} for the more higher  terms $\Lambda_4, \Lambda_5$ and $\Lambda_6$.

We first write the master equation in the form
\begin{equation}
\epsilon^2\frac{d^2\psi}{dx^2}+Q(x)\psi(x)=0,\label{appeq1}
\end{equation}
where $\epsilon$ is the perturbation parameter introduced to keep track of orders in the WKB approximations. We carry out Taylor expansion to $Q(x)$ about the point $x_0$ at which $-Q(x)$ reaches maximum. We have the expansion
\begin{equation}
Q(x)=Q_0+\frac{1}{2}Q_0''z^2+\frac{1}{6}Q_0'''z^3+\frac{1}{24}Q_0^{(4)}z^4+\frac{1}{120}Q_0^{(5)}z^5+\frac{1}{720}Q_0^{(6)}z^6,
\end{equation}
where $z=x-x_0$. Then the master equation \eqref{appeq1} can be rewritten as
\begin{equation}
\epsilon^2 d^2\psi/dz^2+k(-z_0^2+z^2+bz^3+cz^4+dz^5+fz^6)\psi=0,\label{appeq2}
\end{equation}
where
\begin{gather}
k=\frac{1}{2}Q_0'', z_0^2=-2\frac{Q_0}{Q_0''}, b=\frac{1}{3}\frac{Q_0'''}{Q_0''}\\
c=\frac{1}{12}\frac{Q_0^{(4)}}{Q_0''}, d=\frac{1}{60}\frac{Q_0^{(5)}}{Q_0''}, f=\frac{1}{360}\frac{Q_0^{(6)}}{Q_0''}.
\end{gather}
Introducing a new variable $t\propto z/\epsilon^{1/2}$, we could define constants $\nu,\Lambda_2, \Lambda_3$ and  then rescale the parameters $b,c,d,f$ as
\begin{gather}
t=(4k)^{1/4}e^{-i\pi/4}z/\epsilon^{1/2}\\
K\equiv\nu+\frac{1}{2}=-ik^{1/2}z_0^2/2\epsilon-\epsilon\Lambda_2-\epsilon^2\Lambda_3\label{appeq4}\\
\bar{b}=\frac{1}{4}b(4k)^{-1/4}e^{i\pi/4},~~~ \bar{c}=\frac{1}{4}c(4k)^{-1/2}e^{i\pi/2}\\
\bar{d}=\frac{1}{4}d(4k)^{-3/4}e^{3i\pi/4}, ~~~\bar{f}=\frac{1}{4}f(4k)^{-1}e^{i\pi}.
\end{gather}
Then, Eq. \eqref{appeq2} takes the form
\begin{equation}
\frac{d^2\psi}{dt^2}+\left(K-\frac{1}{4}t^2-\epsilon^{1/2}\bar{b}t^3+\epsilon(\Lambda_2-\bar{c}t^4)-\epsilon^{3/2}\bar{d}t^5+
\epsilon^2(\Lambda_3-\bar{f}t^6)\right)\psi=0.\label{appeq3}
\end{equation}
One can refer to \cite{Iyer:1986np} for the further process following \eqref{appeq3}. Here we shall turn our attention to the expressions of correction terms $\Lambda_2, \Lambda_3$
\begin{align}
&\Lambda_2=\frac{1}{2}(3\bar{c}-7\bar{b}^2)+K^2(6\bar{c}-30\bar{b}^2)\\
&\Lambda_3=-K(1155\bar{b}^4-918\bar{b}^2\bar{c}+67\bar{c}^2+190\bar{b}\bar{d}\nonumber\\
&-25\bar{f})-K^3(2820\bar{b}^4-1800\bar{b}^2\bar{c}+68\bar{c}^2
+280\bar{b}\bar{d}-20\bar{f}).
\end{align}
Note that according to Eq. \eqref{appeq4}, we take $\epsilon=1$, $Q(x)=\omega^2-V(x)$, $Q(x_0)=Q_0=\omega^2-V_0$ then we reduces to the 3th-order WKB formula of Eq.\eqref{eq-K} as
\begin{equation}
K=i\frac{\omega^2-V_0}{\sqrt{-2V_0''}}-\Lambda_2(K)-\Lambda_3(K).
\end{equation}
\bibliography{AcusBH}
\end{document}